\documentclass[aps, a4paper,superscriptaddress, nofootinbib, showpacs, twocolumn, 10pt]{revtex4}
\usepackage{amsmath,amssymb,bm,natbib}
\usepackage{epsfig}
\usepackage{graphicx}
\usepackage{slashed}
\newcommand{\beq}{\begin{equation}}
\newcommand{\eeq}{\end{equation}}
\newcommand{\bea}{\begin{eqnarray}}
\newcommand{\eea}{\end{eqnarray}}
\newcommand{\ba}{\begin{align}}
\newcommand{\ea}{\end{align}}
\newcommand{\bfig}{\begin{figure}}
\newcommand{\efig}{\end{figure}}

\newcommand{\D}{\displaystyle}

\newcommand{\gev}{\, \text{GeV}}

\newcommand{\LO}{\, \text{LO}}

\newcommand{\tin}{t_{\text{in}}}
\newcommand{\la}{\langle}
\newcommand{\ra}{\rangle}

\newcommand{\low}{\text{low}}

\newcommand{\up}{\text{up}}
\usepackage{color}

\newcommand{\omnes}{{\cal{O}}}

\newcommand{\fmsq}{\, \text{fm}^2}

\begin{document}

\phantom{}
\vspace*{-17mm}



\title{Precise determination of the low-energy hadronic contribution to the muon $g-2$ from  analyticity and unitarity: An improved analysis}

\author{B.Ananthanarayan}
\affiliation{Centre for High Energy Physics,
Indian Institute of Science, Bangalore 560 012, India}
\author{Irinel Caprini}
\affiliation{Horia Hulubei National Institute for Physics and Nuclear Engineering,
P.O.B. MG-6, 077125 Magurele, Romania}
\author{ Diganta Das}
\affiliation{Physical Research Laboratory,  Navrangpura, Ahmedabad 380 009, India}
\author{I. Sentitemsu Imsong\footnote[2]{Deceased.}}
\affiliation{The Institute of Mathematical Sciences, Taramani, Chennai 600 113, India }

\begin{abstract}
 The two-pion low-energy contribution 
to the anomalous magnetic moment  of the muon, $a_\mu\equiv(g-2)_\mu/2$,  expressed as an integral over the modulus squared of the pion electromagnetic form factor, brings a relatively large contribution to the theoretical error,  since the low accuracy of experimental measurements in this region is amplified by the drastic increase of the integration kernel. We derive stringent constraints on the two-pion contribution by exploiting analyticity and unitarity of the pion electromagnetic form factor. To avoid the poor knowledge of the modulus of this function, we use instead its phase, known with high precision in the elastic region from Roy equations for pion-pion scattering via the Fermi-Watson theorem. Above the  inelastic threshold we adopt a conservative integral condition on the modulus, determined from data and perturbative QCD.  Additional high precision data on the modulus  in the range $0.65-0.71$ GeV, obtained from $e^+e^-$ annihilation and $\tau$-decay experiments, are used to improve the 
predictions  on the modulus 
at lower energies by means of a parametrization-free analytic extrapolation. The results are optimal for a given input and do not depend on the unknown phase of the form factor above the inelastic threshold.
The present work improves a previous analysis based on the same technique, including more experimental data and employing better statistical tools for their treatment. We obtain for the contribution to $a_\mu$ from below 0.63 GeV the value  $(133.258 \pm 0.723)\times 10^{-10}$, which  amounts to a reduction of the theoretical error by about  
 $6 \times 10^{-11}$.
\end{abstract}

\maketitle
\section{Introduction} \label{sec:Intro}
The muon anomalous magnetic moment is one of the most precisely measured observables in particle physics. It can also be predicted by theory with a high accuracy, and is therefore an ideal quantity for testing the standard model and for finding possible deviations from it caused by new physics  \cite{CzMa, DaMa}.   The great interest in muon anomaly is motivated by the present discrepancy  of about 3 to 4 $\sigma$ between theory and experiment. For recent reviews, see Refs. \cite{MiRa, JeNy, Miller, PDG} (see also the bibliography in e.g, \cite{Carloni:2015}).   New generation measurements of muon $g-2$  planned at Fermilab \cite{Fermi} and JPARC \cite{JPARC} are expected to produce results with experimental errors at the level of $16 \times 10^{-11}$, a factor of 4 smaller compared to the  Brookhaven measurement \cite{BNL06}. This therefore requires a precision at the same level also for the theoretical result.

The largest theoretical uncertainties are related to the hadronic contribution to $a_\mu$, which comes mainly from energies at which the confined quarks are strongly  interacting   and the QCD  perturbative  treatment  breaks  down.  The evaluation  of the nonperturbative effects  is usually done by means of dispersion
relations in  conjunction  with  experimental  data. Low energy effective theories and  lattice QCD are also used. Efforts are currently made to increase the precision of these calculations, regarding both the hadronic vacuum polarization and the hadronic light-by-light scattering   (a compilation of recent studies is presented in \cite{Benayounetal}). 

The hadronic vacuum polarization (VP), which is numerically the most significant term, contributes with about $43 \times 10^{-11}$ units to the theoretical error. It is dominated by the two-pion contribution, which brings more than 70\% of the leading-order hadronic contribution. 

The two-pion contribution to the VP is expressed in terms of the modulus squared of the pion electromagnetic form factor. It has been measured in $e^+e^-$-annihilation  experiments by CMD2 \cite{CMD2,CMD204,CMD206}, 
SND \cite{SND}, \emph{BABAR} \cite{BaBar,BaBar1}, KLOE \cite{KLOE1, KLOE2, KLOE3} and BESIII \cite{BESIII},  and from the hadronic decays of the $\tau$ lepton  by CLEO \cite{Anderson:1999ui}, ALEPH \cite{Schael:2005am, Davier:2013sfa},  
  OPAL \cite{Ackerstaff:1998yj} and Belle \cite{Fujikawa:2008ma}. Due to experimental difficulties in the identification of low-energy  pions,  the data  below 0.6 GeV have very large uncertainties, except for \emph{BABAR} and KLOE.
The recent data published by BESIII \cite{BESIII} are restricted to energies above 0.6 GeV. Two new detectors, CMD-3 and SND,  now operating at the VEPP-2000 $e^+e^-$ collider in Novosibirsk, are expected to bring accurate data of greatest interest for the $a_\mu$ evaluation \cite{Achasov:2014xsa, CMD3, SND2}. Preliminary results reported in \cite{Fedotovich:2015kna, EidelmanCD} indicate as goal an accuracy comparable to that of \emph{BABAR} and KLOE experiments. 

The lack of data of sufficient precision at low energies, combined with the fact that the integration kernel exhibits a drastic increase in this region, leads to a relatively large uncertainty of the corresponding contribution to the muon anomaly \cite{Davier:2009, Davier:2011, Teubner}.  
The contribution to $a_\mu$ from below 0.63 GeV, obtained using a fit of the pion form factor in the region near threshold and the direct integration of a compilation of data on $e^+e^-\to\pi^+\pi^-$ cross section between 0.30 and 0.63 GeV,  is quoted in \cite{Davier:2009} with  an error of $13.1 \times 10^{-11}$, while  the direct integration in this range of the \emph{BABAR} data alone  leads to an error of $14.7 \times 10^{-11} $. 

The present large uncertainty of the two-pion contribution to $a_\mu$ from energies below 0.63 GeV  motivated us recently \cite{Ananthanarayan:2013zua} to investigate it theoretically in a framework based on the analyticity and unitarity properties of the pion form factor.  The main idea was to use,  instead of the poorly known modulus,  the phase of the form factor, which is equal by the Fermi-Watson theorem \cite{Fermi:2008zz, Watson} to the $\pi\pi$ scattering  $P$-wave phase shift, which has been calculated with high precision from Chiral Perturbation Theory (ChPT) and Roy equations \cite{ACGL, Caprini:2011ky, GarciaMartin:2011cn}.  Above  the  inelastic  threshold,  where  the Fermi-Watson theorem is no longer valid and the
phase  of the form factor is not known, we have used an integral condition on the form-factor modulus, derived using measurements of the \emph{BABAR} experiment \cite{BaBar,BaBar1}  up to
3 GeV and the asymptotic behavior of the form factor predicted by perturbative QCD \cite{Farrar:1979aw,Lepage:1979zb, Melic:1998qr} above that energy.

 The knowledge of the phase on a part of the unitarity cut and of the modulus on the other part of the cut is  not  sufficient for uniquely predicting the form factor. However,  as shown first in \cite{IC}, from this information one can derive rigorous upper and lower bounds on the modulus below the inelastic threshold, in particular  in the low energy region. To increase the strength of the bounds, we have used as input also several values of the modulus from
the region $0.65-0.71$ GeV,  measured with higher precision by the $e^+e^-$ experiments CMD2 \cite{CMD2},  SND \cite{SND} \emph{BABAR} \cite{BaBar,BaBar1} and KLOE 13 \cite{KLOE3}. 
The  method  amounts to a  parametrization-free  analytic  extrapolation
from higher energies to the low energy region of interest for the improved calculation of the muon anomaly.
It led to a two-pion contribution to $a_\mu$  from the region below 0.63 GeV which agreed with other recent determinations  and had
a smaller uncertainty \cite{Ananthanarayan:2013zua}.  

In the present paper we present an update of the work \cite{Ananthanarayan:2013zua}, improving certain details of the analysis. The main improvement is a proper treatment by Monte Carlo simulations of the statistical errors of the data used as input, which will allow us to attach an uncertainty  to the result at a  precise confidence level (C.L.). 
Also, better tools \cite{MiSc} for combining different predictions accounting for their possible correlations  are used.
In addition to the  $e^+e^-$  data from the region $0.65-0.71$ GeV used as input in \cite{Ananthanarayan:2013zua}, we also consider the   KLOE independent measurements
reported in  \cite{KLOE2} and the very recent data of BESIII \cite{BESIII}.  We include also the data obtained in the same energy region from
$\tau$-lepton decays  by the  
CLEO\cite{Anderson:1999ui}, ALEPH \cite{Schael:2005am, Davier:2013sfa},  OPAL\cite{Ackerstaff:1998yj} and Belle \cite{Fujikawa:2008ma} collaborations. 

The outline of this paper is as follows:  
in Sec. \ref{sec:aim}  we formulate our aim and review the  conditions used as input.  In Sec. \ref{sec:input} we give a detailed description of the 
experimental information used as input and in Sec. \ref{sec:errors} we  describe the Monte Carlo simulation  used for implementing the statistical uncertainties of the input data and the prescription of combining the predictions from different experiments. 
 Section \ref{sec:results} contains our results and
Sec. \ref{sec:conclusion}  a summary and our conclusions.  The paper has two Appendices: in 
Appendix \ref{sec:A} we present the solution of the functional extremal problem formulated  in Sec.  \ref{sec:aim}, which is the mathematical basis of our approach. In
Appendix \ref{sec:B}  we discuss the extraction of the pion form factor from the $e^+e^-$ and $\tau$-decay experiments, giving a short overview of various corrections applied.  

\vspace{0.3cm}
\section{Formalism\label{sec:aim}}
We consider the leading order (LO) two-pion contribution to $a_\mu$, which does not contain the
vacuum polarization effects but includes one-photon final-state radiation (FSR).  We are interested in finding the two-pion contribution to $a_\mu$ from  the interval of energies ranging from $\sqrt{t_\low}$ to $\sqrt{t_\up}$,
which is expressed in terms of the pion electromagnetic form factor $F(t)$ as
\begin{widetext} 
\begin{equation} \label{eq:amu}
a_\mu^{\pi\pi(\gamma), {\text{LO}}} [\sqrt{t_\low}, \sqrt{t_\up}] = \frac{\alpha^2 m_\mu^2}{12 \pi^2}\int_{t_\low}^{t_\up} \frac{dt}{ t}  \, K(t)\, \beta^3_\pi(t) \,   
|F(t)|^2 |F_\omega(t)|^2  \left(1+\frac{\alpha}{\pi}\,\eta_\pi(t)\right).\end{equation}\end{widetext} 
In this relation,  $\beta_\pi(t)=(1-4 m_\pi/t)^{1/2}$ is the two-pion phase space relevant for $e^+ e^-\to \pi^+\pi^-$ annihilation  ($m_\pi$ being the charged pion mass), and
\begin{equation} \label{eq:K}
K(t) = \int_0^1 du(1-u)u^2(t-u+m_\mu^2u^2 )^{-1}
\end{equation}
is the QED kernel function. This function is known to exhibit a drastic increase at low $t$ \cite{CzMa}.

The integrand in (\ref{eq:amu}) contains the pion electromagnetic form factor  $F(t)$ in the isospin limit,  defined by
\beq\label{eq:def}  \langle \pi^+(p')\vert J_\mu^{\rm elm} 
\vert \pi^+(p)\rangle= (p+p')_\mu F(t), ~ t=(p-p')^2,
\eeq
 which is a real analytic function in the $t$ complex plane cut along the real semiaxis $t\ge 4 m_\pi^2$.
The remaining factors in (\ref{eq:amu}) denote corrections not included in the form factor: $F_\omega(t)$ accounts for the isospin violation due to $\rho-\omega$ mixing and is parametrized as
\cite{Leutwyler:2002hm, Hanhart:2012wi}:
\begin{align}\label{eq:rhoomega}
 F_\omega(t)= 1+\epsilon\frac{t}{(m_\omega-i\Gamma_\omega/2)^2 -t}\,,
\end{align}
where $\epsilon=1.9\times 10^{-3}$. Finally, $\eta_\pi(t)$ is the FSR correction, calculated in scalar QED \cite{FSR1, FSR2}.

We are interested in the contribution to (\ref{eq:amu}) of the energies below 0.63 GeV.  For convenience we shall use in what follows the simplified notation 
\beq\label{eq:not}
a_\mu\equiv a_\mu^{\pi\pi(\gamma),\, {\text{LO}}} [2 m_\pi, 0.63 \,\text{GeV}].
\eeq
 We now formulate the conditions on the form factor $F(t)$ adopted as input for constraining the above quantity. Following Ref. \cite{Ananthanarayan:2013zua}, we write these conditions as:
\vskip0.2cm
 1.  Fermi-Watson theorem \cite{Fermi:2008zz, Watson}, which implies:
\beq\label{eq:watson}
{\rm Arg} [F(t+i\epsilon)]=\delta_1^1(t),  \quad\quad 4 m_\pi^2 \le t \le \tin,
\eeq
where $\delta_1^1(t)$ is the phase-shift of the $P$-wave of $\pi\pi$ elastic scattering and 
$\tin$ is the first inelastic threshold.

 2. Normalization  at $t=0$ and the value of the charge radius $\la r^2_\pi \ra$, expressed by:
\beq\label{eq:taylor}
	F(0) = 1, \quad  \quad \left[\frac{dF(t)}{dt}\right]_{t=0} =\D\frac{1}{6} \la r^2_\pi \ra.
\eeq

 3.  An integral condition on the modulus squared above the inelastic threshold, written in the form
\beq\label{eq:L2}
 \D\frac{1}{\pi} \int_{\tin}^{\infty} dt \rho(t) |F(t)|^2 \leq  I,
 \eeq
where $\rho(t)$ is a suitable positive-definite weight, for which the integral converges and an accurate evaluation of $I$ is possible. 

 4.   The value at one  spacelike energy, known from experiment: 
\beq\label{eq:val}
F(t_s)= F_s \pm \epsilon_s, \qquad t_s<0.
\eeq

  5. The modulus at one energy in the elastic region of the timelike axis, known from experiment: 
\beq\label{eq:mod}
|F(t_t)|= F_t \pm \epsilon_t, \qquad  4 m_\pi^2< t_t <\tin.
\eeq

\vskip0.2cm
As in   Ref. \cite{Ananthanarayan:2013zua}, we consider the following functional extremal problem: using as input the conditions 1-5, we derive optimal upper and lower bounds on $|F(t)|$ at all points on the elastic unitarity cut, $4 m_\pi^2<t<\tin$, in particular at energies below 0.63 GeV of interest for the calculation of the quantity (\ref{eq:not}). The solution of the extremal problem and the algorithm for obtaining the bounds are presented for completeness in Appendix \ref{sec:A}. 
In order to operate this machinery, we need high quality phenomenological inputs, which are the subject of the following section.

\section{Phenomenological input  \label{sec:input}}
In this section we briefly describe the input used in the conditions 1-5, expressed in the Eqs. (\ref{eq:watson})-(\ref{eq:mod}) given in the previous section.

The first significant inelastic threshold $\tin$ for the pion form factor is due to the  opening of the $\omega\pi$ channel, {\em i.e.} $\sqrt{\tin}=m_\omega+m_\pi=0.917\,\gev$. 
Below this threshold, we use in  (\ref{eq:watson}) the  
phase shift $\delta_1^1(t)$ from Refs. \cite{ACGL, Caprini:2011ky} and \cite{GarciaMartin:2011cn}, 
which we denote as Bern and Madrid  phase, respectively.

For the charge radius entering (\ref{eq:taylor}) 
 we use the constraint  $\langle r_\pi^2 \rangle \in\, (0.41,\,0.45)\,\fmsq $  derived in \cite{Ananthanarayan:2013dpa}. Since this range was obtained basically from the same constraints as those listed in the previous section, the knowledge of the charge radius plays actually a weak role in further improving the bounds on the modulus in the energy region of interest. However, we keep this condition since we now use a different treatment of the uncertainties compared to our previous analyses.

\begin{figure*}[htb]
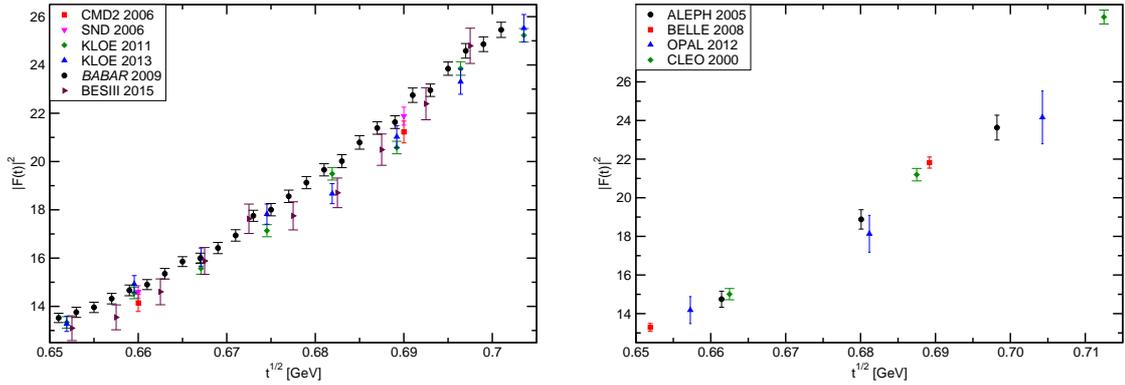
\vspace{0.3cm}
\begin{center}
 \includegraphics[width = 7cm]{fig1_ee.eps}\hspace{0.6cm} \includegraphics[width = 7cm]{fig1_tau.eps}
\caption{Modulus squared $|F(t)|^2$ measured in the region 0.65-0.71 GeV by $e^+e^-$-annihilation  (left) and $\tau$-decay (right) experiments.  \label{fig:epemtau}}\end{center}
\end{figure*}

We have calculated the integral (\ref{eq:L2}) using the \emph{BABAR} data \cite{BaBar} 
from $\tin$ up to $\sqrt{t}=3\, \gev$, smoothly continued with  a constant value for the modulus 
in the range $3\, \gev \leq \sqrt{t} \leq 20 \gev$,  and  a $1/t$ decreasing modulus at higher energies, as predicted by perturbative QCD \cite{Farrar:1979aw,
Lepage:1979zb, Melic:1998qr}. As discussed in detail in Refs. \cite{Ananthanarayan:2013zua, Ananthanarayan:2012tn, Ananthanarayan:2012tt}
this evaluation is expected to overestimate the true value of the integral. 
As in \cite{Ananthanarayan:2013zua} we have adopted the weight $\rho(t)=1/t$, for which 
the contribution of the range above 3 GeV to the integral (\ref{eq:L2}) is 
only of $1\%$. The value of $I$ obtained with this weight is \cite{Ananthanarayan:2013zua}
\beq\label{eq:Ivalue1} 
I=0.578 \pm 0.022, 
\eeq 
where the uncertainty is  due to the \emph{BABAR} experimental errors. In the calculations 
we have used as input for $I$  the central value quoted in Eq. (\ref{eq:Ivalue1}) 
increased by the  error, which leads to the most conservative bounds due to a 
monotonicity property discussed in Appendix A.

For the spacelike  input (\ref{eq:val}) we have used the most recent experimental 
determinations \cite{Horn, Huber}
\bea\label{eq:Huber}	
F(-1.60\,\gev^2)= 0.243 \pm  0.012_{-0.008}^{+0.019}, \nonumber \\ 
F(-2.45\, \gev^2)=  0.167 \pm 0.010_{-0.007}^{+0.013}.
\eea

As shown in \cite{Ananthanarayan:2013zua}, a major role in increasing the strength of the bounds is played by condition (\ref{eq:mod}), with $0.65\gev \leq\sqrt{t_t} 
\leq 0.71\gev$. This energy region was chosen since it is close to  the region of interest and therefore has a stronger effect on improving the bounds than the input from higher energies. 
The $e^+e^-$ data are taken below 0.705 GeV and the $\tau$-decay data below 0.710 GeV, with the exception of
one datum from CLEO that corresponds to an energy of 0.712 GeV.  Since this last datum is at an energy
that is only marginally higher than the upper limit
of the aforementioned energy range, it is included in the analysis.
It is noteworthy that in this region the modulus measured by various experiments exhibits smaller variations than in other 
energy regions and a higher degree of mutual consistency.

 The numbers of experimental points in this range for various experiments, considered in our analysis, are summarized  in Table \ref{tab:1}.  We emphasize that in this region the $e^+e^-$-annihilation and $\tau$-decay experiments are fully consistent, so it is reasonable to use  all the experiments on an equal footing.
\begin{table}[t]
\centering
\renewcommand{\arraystretch}{1.3}
\begin{tabular}{l r }
\hline
Experiment ~~~& Number of points\\
\hline
CMD2 \cite{CMD2} & 2 \\
 SND \cite{SND}& 2  \\
\emph{BABAR}  \cite{BaBar, BaBar1}  & 26 \\
 KLOE 2011 \cite{KLOE2}& 8 \\
 KLOE 2013 \cite{KLOE3}& 8 \\ 
 BESIII  \cite{BESIII}& 10 \\ 
\hline
CLEO \cite{Anderson:1999ui} & 3\\
ALEPH  \cite{Schael:2005am, Davier:2013sfa} & 3\\
OPAL \cite{Ackerstaff:1998yj}&3\\
Belle  \cite{Fujikawa:2008ma} &2\\
\hline
\end{tabular} 
\caption{Number of points in the region $0.65\gev \leq\sqrt{t} 
\leq 0.71\gev$ where the modulus is measured by the $e^+e^-$ annihilation and $\tau$-decay experiments considered in the analysis. \label{tab:1}}
\end{table}
The extraction of the values of timelike modulus $|F(t)|$ from the  cross-section of the process $e^+e^-\to\pi^+\pi^-$  and the spectral function measured in $\tau$-decay experiments implies the application of several corrections, which ensure that the extracted quantity is indeed the form factor $F(t)$ defined in (\ref{eq:def})  in the isospin limit.
Details are given in Appendix \ref{sec:B}. Note that, for OPAL data we have used the rescaled values as presented in \cite{Boito:2012}. For completeness, we show in Fig. \ref{fig:epemtau} the data on modulus squared $|F(t)|^2$ from the $e^+e^-$-annihilation and  $\tau$-decay used as input in our analysis.
It may be observed that the energies at which the form factor measurements are made vary from experiment
to experiment. Therefore, it is not possible to combine the data and bring down the experimental error into the input itself.
\vspace{-0.3cm}
\section{Calculation of $a_\mu$ and its uncertainty \label{sec:errors}}

The algorithm presented in Appendix \ref{sec:A} allows us to obtain rigorous bounds on $|F(t)|$ for $t$ in the region $\sqrt{t}\leq 0.63 \gev$, in terms of the input specified in Sec. \ref{sec:aim}. We recall that the input consists of the phase of the form factor for $t\leq \tin$, the charge radius, one spacelike datum and one timelike modulus from the region $0.65-0.71$ GeV. The lower and upper bounds (\ref{eq:mM}) are given by explicit expressions depending on definite values of the input. In practice, the input values are affected by uncertainties. In order to account for them,  we have generated pseudorandom
numbers for each of the input quantities with  \emph{a priori} given distributions. For the experimental spacelike and timelike data  we have assumed Gaussian distributions with central value as the mean and the quoted errors as the standard deviation. For the spacelike data,  symmetrized errors  have been used in the Gaussian distributions, taking for symmetrization the biggest error \cite{MiSc}. The distributions of the phase and the charge radius, which are calculated from theoretical constraints, were assumed to be uniform.

For a timelike input in 0.65-0.71 GeV region, a specified spacelike input and a selected phase (Bern or Madrid), 
 a large sample ($\sim 10^5$) of sets of inputs have been obtained by randomly drawing one value each from the above distributions.
 For each set of inputs in the sample,  upper and lower bounds on the modulus squared $|F(t)|^2$ have been calculated using 
 the algorithm of Appendix A, at each energy
$\sqrt{t}$ from threshold to 0.63 GeV. All values in between the upper and lower bounds are equally probable. Therefore, 
for a given set of input (which yields one set of upper and lower bounds at each $\sqrt{t}$),
at each low energy point $\sqrt{t}$ from threshold to 0.63 GeV,  a random admissible value
 between the bounds has been generated and used in the integration (\ref{eq:amu}), 
 yielding one value of the quantity $a_\mu$.  
The procedure is repeated 50 times for each set of inputs  
that yields 50 values of the quantity $a_\mu$. With this procedure, at each fixed timelike energy point, we obtain
a large sample ($\sim 10^{6}$) of the quantity $a_\mu$. 
The entire sample is binned to obtain a mean value and 68.3\% confidence level upper and lower bounds at each timelike point.
 
The predictions obtained with input from different timelike energies were then combined into an average result for each experiment. 
The procedure of obtaining the average of several measurements in principle requires the knowledge of the correlations between the different values. 
Since these are not known\footnote{One can use as a first indication the bin-to-bin correlations of the input data on the modulus, which can be extracted in some cases from the published  covariance matrices.},   we applied the averaging  prescription proposed in \cite{MiSc}, where the effective size of the correlations is estimated from the data themselves. As discussed in \cite{MiSc}, the most robust average of a set of $n$ measurements $a_i$ is the weighted average
\beq\label{eq:av}
\bar{a}=\sum_{i=1}^{n} w_i a_i, \quad w_i=\frac{1/\delta a_i^2}{\sum_{j=1}^{n}1/\delta a_j^2},
\eeq
where $\delta a_i$ is the error of $a_i$.

For the best estimation of the error in the case of unknown correlations,  the prescription proposed in \cite{MiSc} is to define a function $\chi^2(f)$  
\beq\label{eq:chisq}
\chi^2(f) = \sum_{i,j=1}^{n} (a_i- \bar{a})(C(f)^{-1})_{ij}(a_j-\bar{a}) 
\eeq
in terms of the covariance matrix $C(f)$ with elements
\begin{equation}\label{eq:cov}
  C_{ij}=\begin{cases}
    \delta a_i \delta a_i & \quad \quad \text{if $i=j$},\\
    f  \delta a_i \delta a_j &\quad \quad \text{if $i\neq j$}.
  \end{cases}\
\end{equation}
The parameter  $f$ denotes the fraction of the maximum possible correlation: for $f=0$ the measurements are treated as uncorrelated, for  $f=1$ as fully (100\%) correlated.

If $\chi^2(0)<n-1$, the data might indicate the existence of a positive correlation. The prescription proposed in \cite{MiSc} is to increase $f$ until $\chi^2(f)=n-1$.
With the solution $f$ of this equation, the standard deviation $\sigma (\bar a)$ of $\bar a$ is determined from the variance \cite{MiSc}
\beq\label{eq:sigma}
\sigma^2 (\bar a)=\left (\sum_{i,j=1}^n (C(f)^{-1})_{ij}\right)^{\!\!-1}.
\eeq 
On the other hand, if one obtains $\chi^2(0)>n-1$, this is an  indication that the individual errors are underestimated.  If the ratio $\chi^2(0)/(n-1)$ is not very far from 1, the procedure suggested in \cite{MiSc, PDG} is to rescale the variance $\sigma^2 (\bar a)$ calculated with (\ref{eq:sigma})  by the factor $\chi^2(0)/(n-1)$. In our work, this kind of procedure was applied first for combining the results obtained with different measurements by each experiment. Then  the predictions of various experiments were combined leading to a global average. 

\begin{figure}[thb]\vspace{0.5cm}
\begin{center}
 \includegraphics[width = 8cm]{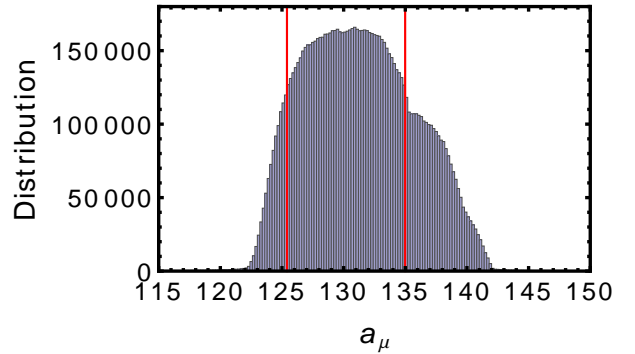}
\caption{Distribution of $a_\mu$ values obtained from the Monte Carlo sample of pseudodata, without input modulus. The vertical lines delimitate the region of 68.3\% C.L. \label{fig:fig1}}
\end{center}
\vspace{0.3cm}
\end{figure}

\section{Results \label{sec:results}}

It is instructive to first give the value obtained without using as input the measurements of the timelike modulus.
In Fig. \ref{fig:fig1} we show the distribution of the $a_\mu$ sample, obtained using as input the Bern phase and the first spacelike point from (\ref{eq:val}). It may be readily seen that the distribution is not fully symmetrical, as it should be for a Gaussian distribution. From this distribution, by applying the 68.3\% criterion we obtained for $a_\mu$ the value $(130.865^{+4.124}_{-5.460}) \times 10^{-10}$, and Madrid phase gives a similar result, $(131.933 ^{+3.438 }_{-5.922}) \times 10^{-10}$. Since these results are not statistically independent, the most conservative procedure is to take the simple average of the central values and of the uncertainties. This gives
\beq\label{eq:nomod}
 a_\mu^{\pi\pi, \LO}\,[2m_\pi ,\, 0.63 \gev]=(131.399 ^{+3.780 }_{-5.691}) \times 10^{-10}.
\eeq
The large error shows that the constraining power of the phase and the spacelike data is rather low.

\begin{figure}[thb]\vspace{0.5cm}
\begin{center}
 \includegraphics[width = 8cm]{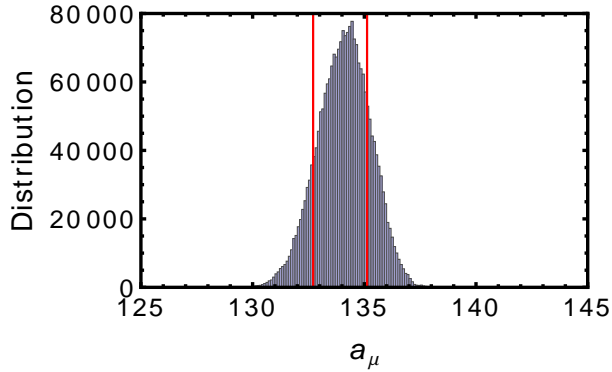}
\caption{Distribution of a typical $a_\mu$ sample obtained from the Monte Carlo simulation of pseudodata, with an input  modulus measured by \emph{BABAR} in the region $0.65-0.71$ GeV. Details of the input are given in the text. The vertical lines delimitate the region of 68.3\% C.L. \label{fig:fig2}}
\end{center}
\vspace{0.3cm}
\end{figure}

By including as input the modulus measured at one energy from the region $0.65 - 0.71$ GeV, the determination (\ref{eq:nomod}) is considerably improved. In Fig. \ref{fig:fig2} we show for illustration the distribution of the $a_\mu$ sample, obtained using as input the Bern phase, the highest energy \emph{BABAR} point shown in Fig. \ref{fig:epemtau} and the first spacelike point from (\ref{eq:val}). The distribution is much narrower than that shown in Fig. \ref{fig:fig1} and more symmetrical, allowing the extraction of a smaller standard deviation by means of the 68.3\% C.L. criterion. 

Similar distributions of $a_\mu$  have been obtained for all the values of the input  modulus measured in the region $0.65-0.71$ GeV, shown in Fig. \ref{fig:epemtau}. The procedure was applied for each of the input phases, Bern and Madrid.  The calculation was performed using as input  each of the two spacelike values (\ref{eq:val}) and the best prediction was retained. 

\begin{figure}[htb]\vspace{0.4cm}
\begin{center}
 \includegraphics[width = 9cm]{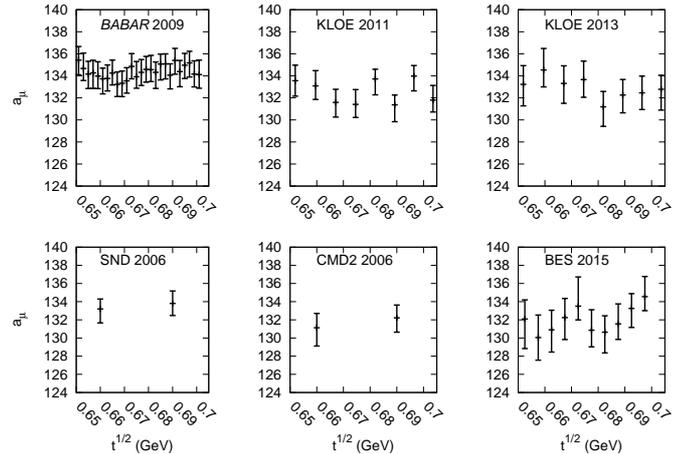}
\caption{Allowed intervals at 68.3\% C.L. for the quantity
$a_\mu \equiv a_\mu^{\pi\pi(\gamma),\, {\text{LO}}} [2 m_\pi, \,0.63\, \text{GeV}]\times 10^{10}$, as a function of the energy in the region $(0.65 - 0.71)\, \text{GeV}$ where the timelike modulus used as input was measured in $e^+e^-$ experiments. \label{fig:fig2e}}\end{center}
\vspace{0.cm}
\end{figure}

\begin{figure}[htb]\vspace{0.4cm}
\begin{center}
 \includegraphics[width = 9cm]{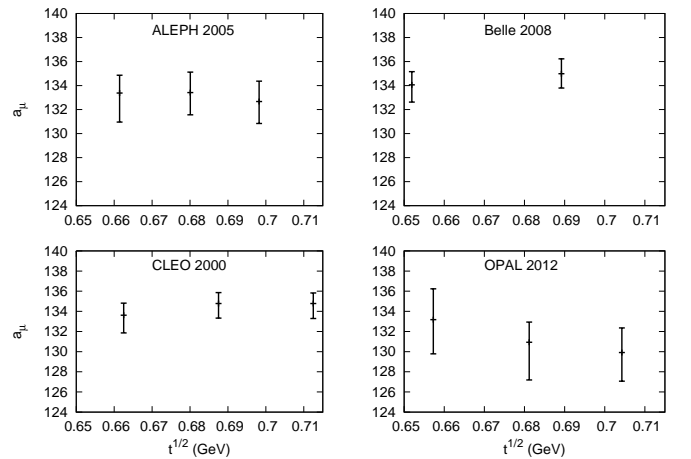}
\caption{Allowed intervals at 68.3\% C.L. for the quantity $a_\mu \equiv a_\mu^{\pi\pi(\gamma),\, {\text{LO}}} [2 m_\pi, \,0.63 \,\text{GeV}]\times 10^{10}$, as a function of the energy in the region $0.65 - 0.71\, \text{GeV}$  where the timelike modulus  used as input was measured in  $\tau$ experiments.  \label{fig:fig2t}}\end{center}
\vspace{0.cm}
\end{figure}

In Figs. \ref{fig:fig2e} and \ref{fig:fig2t} we show the 68.3\% C.L. intervals of $a_\mu$ obtained from the Monte Carlo simulation described in the previous section,  for all the timelike points used as input from the $e^+e^-$  and  $\tau$ experiments.  The results have been obtained using as input the Bern phase \cite{ACGL, Caprini:2011ky}. The Madrid phase leads to similar results.  The bounds obtained with various values of the timelike modulus  reflect the quality of data shown in
Figs. \ref{fig:epemtau}, ranging between the most accurate, \emph{BABAR}, and those with the largest errors, OPAL.

We have then applied the averaging procedure described in the previous section, for combining the predictions available from different measurements of each experiment. The average was obtained using the robust prescription (\ref{eq:av}). For estimating the error, we have computed $\chi^2(f)$ defined in (\ref{eq:chisq}) and compared it with the number of degrees of freedom, $n-1$, where $n$ is the number of points in each panel of Figs. \ref{fig:fig2e} and \ref{fig:fig2t}. It turned out that in all cases  the ratio $\chi^2(0)/(n-1)$ was less than 1  and increased for a positive correlation, reaching unity for $f$ in general in the range $0.40-0.70$.
  
\begin{figure}[htb]\vspace{0.5cm}
\begin{center}
 \includegraphics[width = 6.7cm]{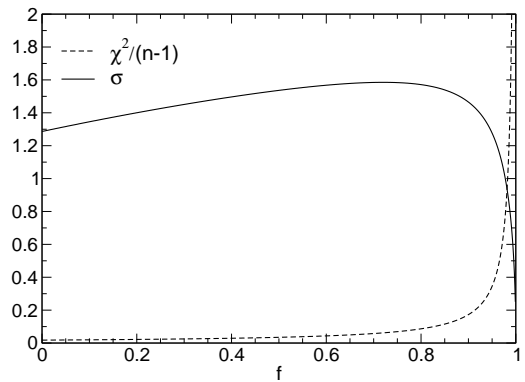}
\caption{Dependence on  $f$ of the ratio $\chi^2(f)/(n-1)$ and the standard deviation, $\sigma\equiv\sqrt{\sigma^2(f)}$ for the timelike data measured by CMD2 and Madrid phase. The equality $\chi^2(f)/(n-1)=1$ holds for large values of $f$.  \label{fig:fig3a}}\end{center}
\vspace{0.2cm}
\end{figure}

Some pathologies were encountered however in a few cases.  One type of pathology is illustrated  in Fig. \ref{fig:fig3a}, where we show the dependence on $f$ of the ratio $\chi^2(f)/(n-1)$ and of the standard deviation for the input from CMD2 and Madrid phase. In this case, the ratio becomes 1 only for values of $f$ close to 1, where the variance $\sigma^2(f)$ calculated according to  (\ref{eq:sigma}) starts to decrease\footnote{One can show that in all cases when the individual errors are different, $\sigma$ exhibits a decrease above a certain $f$ and vanishes for $f=1$. In the particular case of equal errors, the variance grows linearly  with $f$, as discussed in \cite{MiSc}.}. This happens because the individual values are much closer than expected from the ascribed errors. As discussed in \cite{MiSc}, in such cases the averaging cannot reduce the overall error, as the blind application of the prescription would indicate. Therefore, for this case we adopt the modified prescription of taking the 
maximum variance for $f$ in the range (0, 1). The value of $\chi^2$ corresponding to this $f$ is smaller than 1, which is due to the fact that the individual values to be averaged are very close. We encountered a similar situation with the data from ALEPH and both phases. In these cases, the combined error is not much less than the individual errors entering the combination.

\begin{figure}[htb]\vspace{0.5cm}
\begin{center}
 \includegraphics[width = 6.7cm]{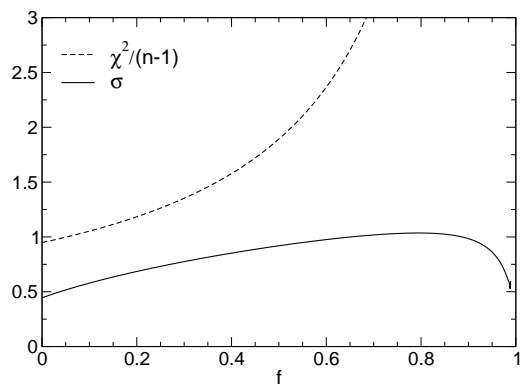}
\caption{Dependence on $f$ of the ratio $\chi^2(f)/(n-1)$ and the standard deviation, $\sigma\equiv\sqrt{\sigma^2(f)}$ for the timelike data measured by KLOE 11 and Madrid phase. The equality $\chi^2(f)/(n-1)=1$ holds for small values of $f$.  \label{fig:fig3b}}\end{center}
\vspace{0.2cm}
\end{figure}

A different type of pathology was encountered with KLOE 11 data: in this case, for both phases, the individual values are rather different and their errors are rather small. As a consequence,  $\chi^2(f)/(n-1)$ becomes 1 for $f$ close to 0. However, the corresponding variance (\ref{eq:sigma}) turns out to be much smaller than estimated from the spread of the individual values. Since these values are based on measurements of the modulus at different energies by the same experiment, the differences among them indicate a problem with the data and an error reduction by their combination is not reliable. Therefore, as a conservative error, we adopted in this case too the maximum variance for $f$ in the range (0, 1), whose magnitude is comparable with those of the individual errors. We illustrate this case  in Fig. \ref{fig:fig3b}, where we show the dependence on $f$ of the ratio $\chi^2(f)/(n-1)$ and of the standard deviation for the input from KLOE 11 and Madrid phase.

\begin{figure}[htb]\vspace{0.5cm}
\begin{center}
 \includegraphics[width = 6.7cm]{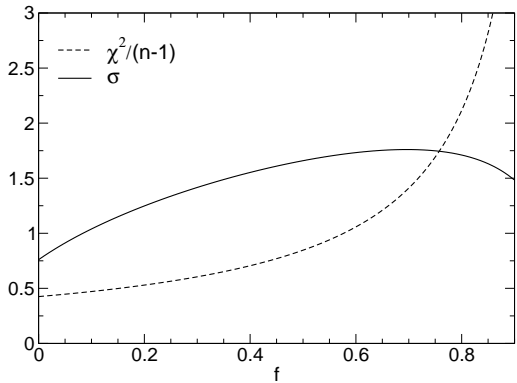}
\caption{Dependence on $f$ of the ratio $\chi^2(f)/(n-1)$ and the standard deviation, $\sigma\equiv\sqrt{\sigma^2(f)}$ for the timelike data measured by BESIII and Madrid phase. The error is obtained with $f$ determined from the equation $\chi^2(f)/(n-1)=1$. \label{fig:fig3c}}\end{center}
\vspace{0.2cm}
\end{figure}

Except these special cases, the standard deviation was calculated using (\ref{eq:sigma}), with the covariance matrix (\ref{eq:cov}) corresponding to the fraction $f$  determined from the equation $\chi^2(f)=n-1$. A typical situation is shown in Fig. \ref{fig:fig3c}, where we show the dependence on $f$ of the ratio $\chi^2(f)/(n-1)$ and of the standard deviation for the input from BESIII and Madrid phase.

In Table \ref{table:amu}, we present the results of the average procedure for all the $e^+e^-$ and $\tau$ experiments.
For completeness, we give the results obtained separately with the Bern and the Madrid phase. 

 \begin{table}\vspace{0.3cm}
 \begin{tabular}{l c c }\hline \hline
 & ~~Bern phase & ~~~ Madrid phase\\\hline
 CMD2 06 & ~ $ 131.804 \pm 1.563$  & ~~ $ 131.396 \pm  1.585 $  \\
 SND 06 & ~ $ 133.535 \pm 1.371  $ &  ~~ $ 133.102 \pm 1.306 $  \\
 \emph{BABAR} 09  & ~ $ 134.338 \pm 0.939 $  & ~~ $134.086  \pm 0.862 $ \\
 KLOE 11 & ~ $ 132.560 \pm 1.220 $  & ~~ $ 132.017 \pm 1.035 $   \\
 KLOE 13 & ~ $  132.864 \pm 1.413 $ & ~~ $ 132.343 \pm 1.224 $  \\
 BESIII 15 & ~ $ 131.958 \pm 1.725 $ & ~~ $ 132.753 \pm 1.719 $  \\\hline
 CLEO 00 & ~ $134.478 \pm 1.389 $  & ~~ $ 133.897 \pm 1.183 $  \\
 ALEPH 05 & ~ $  133.114 \pm 1.703 $  & ~~ $ 132.298 \pm 1.783 $  \\
 Belle 05 & ~ $  134.588 \pm 1.227 $  & ~~ $ 134.280 \pm 1.136 $ \\
 OPAL 12 & ~ $ 131.176\pm 2.803 $ & ~~ $ 129.910 \pm 2.970 $ \\
\hline\hline
 \end{tabular}\caption{Central values and  68.3\% C.L. standard deviations for the quantity  $a_\mu^{\pi\pi(\gamma),\, {\text{LO}}} [2 m_\pi, \,0.63 \,\text{GeV}] \times 10^{10}$, 
 obtained by averaging the results shown in Figs. \ref{fig:fig2e} and \ref{fig:fig2t} for each experiment. \label{table:amu}}
 \end{table}

 The last step is to combine  the individual values obtained with  measurements by different  experiments. The correlation between these values is difficult to assess \emph{a priori}. There is of course a consistent common information going as input into all these determinations. However, the most important input, which has the crucial role in error reduction, is the modulus of the form factor in the region $0.65-0.71$ GeV measured by different experiments, which makes the difference between the values given in Table \ref{table:amu}. Some correlation might exist also between these measurements, but there is no consensus in the views on their treatment \cite{Davier:2011, Teubner}. We therefore applied the same averaging procedure \cite{MiSc} suitable for cases when the correlations are not known.

The data from $e^+e^-$-annihilation and $\tau$-decay experiments are consistent in the region $0.65-0.71$ GeV, so  the results from all the 10 experiments can be combined into a single average. 
The ratio $\chi^2(0)/(N-1)$, where $N=10$, turned out to be smaller than 1, which indicates a positive correlation between the predictions of various experiments. By applying the prescription given in \cite{MiSc}, the correlation was found to be $f=0.4$ for Bern phase and  $f=0.3$ for Madrid phase, leading to the values of $a_\mu$ equal to $(133.425 \pm 0.793) \times 10^{-10}$ and $(133.092 \pm 0.653)\times 10^{-10}$, respectively\footnote{Assuming the values not correlated, one would obtain considerably smaller errors, 0.437 and   0.402, respectively.}. Taking the simple average of the central values and errors obtained with the two phases, which are not statistically independent, we obtain
the conservative final estimate\footnote{The separate combination of the results obtained with data from $e^+e^-$ and $\tau$-decay experiments leads to the values $(133.018 \pm 0.766) \times 10^{-10}$ and $( 133.785 \pm 0.993)\times 10^{-10}$, respectively.}
 \beq\label{eq:average}
a_\mu^{\pi\pi(\gamma),\, {\text{LO}}} [2 m_\pi, \,0.63 \,\text{GeV}] =(133.258 \pm 0.723 )\times 10^{-10}.
  \eeq
This result is consistent with our previous result reported in Ref. \cite{Ananthanarayan:2013zua} and has a slightly smaller error. We emphasize that in \cite{Ananthanarayan:2013zua} the prediction based on the present formalism was  combined also with the direct integration of the cross section measured by \emph{BABAR}  at energies below 0.63 GeV, while in this work we do not use data from low energies.

\section{Discussions and Conclusion \label{sec:conclusion}}


In this work, we have studied the two-pion contribution from energies below 0.63 GeV to the muon $g-2$,
 by exploiting analyticity and unitarity of the pion pion electromagnetic form factor. The motivation of the work is the relatively large error (of about $13.1 \times 10^{-11}$, see Ref.  \cite{Davier:2009})  of this contribution obtained by direct data integration, explained by experimental difficulties in identifying pion pairs at low energies  and the behavior of the QED kernel $K(t)$ in the integral (\ref{eq:amu}) expressing $a_\mu$ in terms of the pion form factor modulus. 

The main idea of our approach was to use, instead of the modulus, the
phase of the pion form factor in the elastic region, equal by Fermi-Watson theorem to the phase shift of the P-wave $\pi\pi$ amplitude, known with precision  from the solution of Roy equations  \cite{ACGL, Caprini:2011ky, GarciaMartin:2011cn}. We have also used a conservative integral constraint on the modulus above the inelastic threshold, derived from \emph{BABAR} data  \cite{BaBar} and perturbative QCD \cite{Farrar:1979aw, Lepage:1979zb, Melic:1998qr}, and two precise measurements of the form factor at spacelike values of the momentum transfer \cite{Horn, Huber}.

A significant contribution to the final precision is brought by the inclusion in the input of several measurements of the modulus of the form factor at higher energies, from the $e^+e^-$-annihilation experiments CMD2, SND, \emph{BABAR}, KLOE 11, KLOE 13 and BESIII,   and the $\tau$-decay experiments CLEO, ALEPH, OPAL and Belle. In practice we considered the energy region $0.65-0.71$ GeV,  where the modulus is measured with a better accuracy, and which is close enough to the low-energy region of interest such as to have a significant constraining power. From this input, using techniques of functional optimization theory, 
we derived rigorous constraints on the contribution to $a_\mu$ of the energies below 0.63 GeV, where the experimental data are poor. We emphasize that the formalism   exploits in an optimal way the input information and requires no parametrization of the pion form factor. Furthermore, the results  do not depend on the unknown phase of the pion form factor above the inelastic threshold. 

The present analysis supersedes our previous work \cite{Ananthanarayan:2013zua}, where the same mathematical formalism was applied with data from only 4  $e^+e^-$ experiments  (CMD2, SND, \emph{BABAR} and KLOE 13). We included now as input data in the region $0.65-0.71$ GeV from 2 additional $e^+e^-$ experiments,  KLOE 2011  and BESIII 2016, and the measurements in the same region reported by 4  $\tau$-decays experiments. The analysis has been also improved by a proper treatment  with statistical tools of the uncertainties and the correlations between the input data.  For each timelike input from the region  $0.65-0.71$ GeV of a given experiment, we have evaluated a range for $a_\mu$ at a 68.3\% confidence level. The results obtained with a definite input from the region $0.65-0.71$ GeV have been then combined using a statistical prescription suitable for cases when the  correlations among the individual measurements are not precisely known \cite{PDG, MiSc}. The combination of the values obtained with data 
from 
various 
experiments was done using the same prescription, which defines a robust central value  and leads to a conservative error. By this procedure we have increased the reliability of our determination. 

The final outcome of our
analysis is expressed in Eq. (\ref{eq:average}).  Our result is consistent with and more precise than the previous result reported in \cite{Ananthanarayan:2013zua}.  It has an  uncertainty smaller by about $6 \times 10^{-11}$ than the direct integration of the cross section below 0.63 GeV  \cite{Davier:2009}.

Our work demonstrates that very general methods of unitarity and analyticity
can be combined with high precision data from one sector to obtain stringent
constraints in another sector. Using in addition suitable
statistical methods to account for the uncertainties and the correlations of the input data, we obtained a significant
improvement of the low-energy two-pion contribution to $a_\mu$.  While the central
value continues to remain stable, which in itself is a remarkable result,
the fact that it has been possible to lower the uncertainty in this region
by nearly a factor of two makes the results in this paper to be of
significance. Until the accurate data  expected  from the CMD-3 and SND experiments at the VEPP-2000 $e^+ e^−$ collider in Novosibirsk become available, our result represents the most precise and robust determination of the low-energy hadronic contribution to muon $g-2$.

\subsection*{ACKNOWLEDGEMENTS} The authors would like to dedicate this work to the loving memory
of their friend and collaborator I. Sentitemsu Imsong who sadly passed away
on October 30, 2015 while this work was in progress. 
We thank S. Eidelman, F. Jegerlehner, B. Malaescu, A. Nyffeler and M. Zhang
for useful discussions in various stages of the work, and D. Boito for valuable information on the updated OPAL data.  I.C. acknowledges support from CNCS-UEFISCDI, Contract  
Idei-PCE 121/2011 and from ANCS, Contract PN 16 42 01 01/2016. The work of D.D.  was supported by Deutsche Forschungsgemeinschaft Research Unit 
FOR 1873 “Quark Flavour Physics and Effective Theories”.  


\appendix

\section{SOLUTION OF THE EXTREMAL PROBLEM FORMULATED IN SEC. \ref{sec:aim}}\label{sec:A}
We must  find optimal upper and lower bounds on $|F(t)|$ on the elastic unitarity cut, $t_+<t<\tin$  for $F(t)\in {\cal C}$, where  ${\cal C}$ is the class of functions real analytic in the $t$-plane cut along the real axis for $t\ge t_+$,  which satisfy the  conditions 1-5 given in Sec. \ref{sec:aim}.
 By means of a proof presented for the first time in Ref. \cite{IC}, this problem can be reduced to a standard analytic interpolation problem  \cite{Duren} (also known as a Meiman problem \cite{Meiman}).

The first step of the proof is to define  the Omn\`es function 
\beq	\label{eq:omnes}
 \omnes(t) = \exp \left(\D\frac {t} {\pi} \int^{\infty}_{4 m_\pi^2} dt' 
\D\frac{\delta (t^\prime)} {t^\prime (t^\prime -t)}\right),
\eeq
where $\delta(t)$ is equal to $\delta_1^1(t)$  at
$t\le \tin$ and is an arbitrary smooth (Lipschitz continuous) function above $\tin$,
which approaches asymptotically $\pi$. 

It follows that  the function $h(t)$ defined by
\beq\label{eq:h}
F(t)=\omnes(t) h(t)
\eeq
is real on the real axis below $\tin$, therefore it is analytic in the $t$-plane cut only for $t>\tin$. 
In terms of $h(t)$, the equality (\ref{eq:L2}) writes as
\beq\label{eq:hL2}
\D\frac{1}{\pi} \int_{\tin}^{\infty} dt\, 
\rho(t) |\omnes(t)|^2 |h(t)|^2 \leq  I.
\eeq
 This relation can be written in a canonical form if we perform the conformal transformation
 \beq\label{eq:ztin}
 \tilde z(t) = \frac{\sqrt{\tin} - \sqrt {\tin -t}} {\sqrt{\tin} + \sqrt {\tin -t}}
 \eeq
and express the factors multiplying  $|h(t)|^2$ in terms of an outer function, 
{\it i.e.} a function analytic and without zeros in 
the unit disk $|z|<1$. In practice, it is convenient to construct it as a product of two outer functions  \cite{IC, Abbas:2010EPJA}: the first one, denoted as $w(z)$, has the modulus equal to  
$\sqrt{\rho(t)\, |{\rm d}t/ {\rm d} \tilde z(t)|}$. 
For the choice $\rho(t)=1/t$, it is given by the simple expression 
\beq\label{eq:outerfinal0}
w(z)=\sqrt{\frac{1-z}{1+z}}.
\eeq 
The second outer function, denoted as $\omega(z)$,  
  has the modulus equal to  $|\omnes(t)|$, and can be calculated by the integral representation
\beq\label{eq:omega}
 \omega(z) =  \exp \left(\D\frac {\sqrt {\tin - \tilde t(z)}} {\pi} \int^{\infty}_{\tin}  \D\frac {\ln |\omnes(t^\prime)|\, {\rm d}t^\prime}
 {\sqrt {t^\prime - \tin} (t^\prime -\tilde t(z))} \right).
\eeq 

If we define the function $g(z)$ by
\beq	\label{eq:gF11}
 g(z) = w(z)\, \omega(z) \,h(\tilde t(z)),
\eeq 
where  $\tilde t(z)$ is the inverse of $z = \tilde z(t)$  defined in
Eq.(\ref{eq:ztin}), the condition (\ref{eq:hL2}) can be written with no loss of information as
 \beq\label{eq:gI1}
 \frac{1}{2 \pi} \int^{2\pi}_{0} {\rm d} \theta |g(\zeta)|^2 \leq I, \qquad \zeta=e^{i\theta}.
 \eeq
As shown  in the analytic interpolation theory \cite{Meiman, Duren}, 
this condition leads to rigorous correlations 
among the values of the analytic function  $g(z)$ and its derivatives 
at points inside the holomorphy domain, $|z|<1$.  In particular, one can show (for a proof and earlier references see Ref. \cite{Abbas:2010EPJA}) that (\ref{eq:gI1}) implies the positivity condition
\beq\label{eq:posit}
{\cal D}\ge 0
\eeq
 of the determinant ${\cal D}$ defined as
\beq\label{eq:det}
{\cal D}=\left|
\begin{array}{c c c c c c}
\bar{I} & \bar{\xi}_{1} & \bar{\xi}_{2} & \ldots & \bar{\xi}_{N}\\	
	\bar{\xi}_{1} & \D \frac{z^{2K}_{1}}{1-z^{2}_1} & \D
\frac{(z_1z_2)^K}{1-z_1z_2} & 	\ldots & \D \frac{(z_1z_N)^K}{1-z_1z_N} \\
	\bar{\xi}_{2} & \D \frac{(z_1 z_2)^{K}}{1-z_1 z_2} & 
\D \frac{(z_2)^{2K}}{1-z_2^2} &  \ldots & \D \frac{(z_2z_N)^K}{1-z_2z_N} \\
	\vdots & \vdots & \vdots & \vdots &  \vdots \\
	\bar{\xi}_N & \D \frac{(z_1 z_N)^K}{1-z_1 z_N} & 
\D \frac{(z_2 z_N)^K}{1-z_2 z_N} & \ldots & \D \frac{z_N^{2K}}{1-z_N^2} \\
	\end{array}\right|,
\eeq
 in terms of the quantities
\beq\label{eq:barxi}
 \bar{I} = I - \sum_{k = 0}^{K-1} g_k^2, \quad  \quad \bar{\xi}_n =\xi_n - \sum_{k=0}^{K-1}g_k z_n^k.
\eeq where:
\bea\label{eq:values} 
g_k&=&\left[\D \frac{1}{k!} \frac{ d^{k}g(z)}{dz^k}\right]_{z=0}, \quad
0\leq k\leq K-1, \nonumber\\
\xi_n&=& g(z_n), \quad  \quad1\leq n \leq N.\eea
The inequality (\ref{eq:posit}) defines an allowed domain for the real values $g(z_n)$ of the function at $N$ real points $z_n\in(-1,1)$, and the first $K$ derivatives $g_k$ at $z=0$.  
 In our application we consider $K=2$, noting that the coefficients $g_0$ and $g_1$ entering (\ref{eq:values} ) depend on the charge radius $\langle r^2_\pi \rangle$ defined in (\ref{eq:taylor}). We further take $N=3$, choosing two points as input, $t_1=t_s$ and $t_2=t_t$  from the conditions (\ref{eq:val}) and (\ref{eq:mod}),
 while  $t_3$  is an arbitrary point below $\tin$. For $t_1<0$  we have from Eqs.
(\ref{eq:h}) and (\ref{eq:gF11})
\beq\label{xin}
g(z_1)=w(z_1)\, \omega(z_1) \,F(t_1) /\omnes(t_1), \quad z_1=\tilde z(t_1).
\eeq
while for $t_n$, $n=2,3$ we have
\beq\label{eq:gFn}
 g(z_n) = w(z_n)\, \omega(z_n) \,|F(t_n)| /|\omnes(t_n)|,  \quad  z_n=\tilde z(t_n),
\eeq
 where the modulus $|\omnes(t)|$ of the Omn\`es function is obtained from (\ref{eq:omnes}) by the principal value (PV) Cauchy integral
\beq	\label{eq:modomnes}
 |\omnes(t)| = \exp \left(\frac {t} {\pi} \text{\rm PV} \int^{\infty}_{4 m_\pi^2} dt' 
\D\frac{\delta (t^\prime)} {t^\prime (t^\prime -t)}\right).
\eeq
The condition (\ref{eq:posit}) provides the solution of the extremal problem formulated above: indeed, it can be written as a quadratic inequality for the unknown modulus $|F(t_3)|$. Recalling that $t_3$ is an arbitrary point in the elastic region, we obtain from (\ref{eq:posit}) the rigorous condition
\beq\label{eq:mM}
m\leq |F(t)| \leq M, \quad\quad t<\tin
\eeq
where the bounds $m$ and $M$ are calculable in terms of known quantities. 

One can prove \cite{IC,Abbas:2010EPJA}, that the bounds  are optimal and their values do not depend on the unknown phase of the form factor above the inelastic threshold $\tin$. Furthermore, for a fixed weight $\rho(t)$ in (\ref{eq:L2}),  the  bounds  become stronger/weaker when the 
value of the r.h.s $I$ is decreased/increased, respectively. These properties make the formalism model independent and robust against the uncertainties from the high energy region. 

In the procedure above we assumed only one input value on the modulus at a timelike energy. Actually, the general inequality (\ref{eq:posit}) allows the simultaneous inclusion of more input values, which are expected to lead to stronger bounds (\ref{eq:mM}) on the modulus below 0.63 GeV. However, it turns out that when the number of input points is increased the procedure becomes quickly difficult  numerically. It is more convenient to use as input only one  modulus at a time and  combine then the results taking into account the possible correlations between them. 

\section{FORM FACTOR EXTRACTION FROM DATA \label{sec:B}}
In this appendix, we briefly discuss the extraction of the modulus of the form factor and
the various corrections which must be taken into account while extracting the form factor
from the data from $e^+e^-$ annihilation and $\tau$ decay experiments. There is a vast literature on this subject (see  \cite{Davier:2010}-\cite{Jegerlehner:2011ti} and references therein).

In the case of $e^+e^-$ experiments,  the values of timelike form factor
is extracted from the measured cross-section of $e^+e^-\to\pi^+\pi^- (\gamma)$. 
 Several experimental collaborations (CMD2, SND, \emph{BABAR}, KLOE) include the vacuum polarization (VP) into the definition of the pion form factor. Therefore, to obtain $|F(t)|$   we remove VP  from  the modulus quoted in Refs. \cite{CMD2, SND, BaBar,  BaBar1,   KLOE2, KLOE3}. Equivalently, we use
\begin{equation}
|F(t)|^2=\frac{3t}{\alpha^2 \pi \beta_\pi(t)^3} \frac{\sigma^0_{\pi\pi(\gamma)}(t)}{1+\frac{\alpha}{\pi}\,\eta_\pi(t)},
\end{equation}
where  $\sigma^0_{\pi\pi(\gamma)}$ is the undressed cross-section obtained by removing VP and the
$\rho-\omega$ interference factor from the measured cross-section, and $\eta_\pi(t)$ is the FSR factor discussed below Eq. (\ref{eq:amu}).

The  usefulness of $\tau$ decays for the calculation of the hadronic contribution to $a_\mu$  is based on conserved vector current (CVC) hypothesis, which implies the equality  $F^-(t)=F(t)$ of the charged form factor $F^-(t)$ relevant in $\tau^-\to\pi^-\pi^0\nu_\tau$ decay and the form factor defined in (\ref{eq:def}).
For a long time, $\tau$ hadronic decays offered the most precise data for the calculation of the hadronic contribution to $a_\mu$. The increasing precision of the $e^+e^-$ experiments, in particular based on radiative return method, now make the two 
approaches comparable.

 The $\tau$-decay data are given in 
terms of invariant hadronic mass squared distribution.
The modulus of the pion form factor is extracted from the $\pi\pi$ distribution using the relation
\bea\label{eq:FFtau}
|F^-(t)|^2& =& \frac{2m_\tau^2}{|V_{ud}|^2} \frac{1}{S_{EW}} \left(1-\frac{t}{m_\tau^2}\right)^{-2}\left(1+\frac{2t}{m_\tau^2}\right)^{-1} \nonumber \\
 &\times & \frac{\cal{B}_{\pi\pi}}{\cal{B}}  \left(\frac{1}{N_{\pi\pi}}\frac{dN_{\pi\pi}}{dt}\right)
  \frac{1}{\beta_{-}^3(t)}\frac{1}{G_{EM}},
 \eea 
where  $\cal{B}_{\pi\pi}$ is the branching fraction for the 
$\tau$ decay into a dipion pair, $\cal{B}$ denotes the electron branching fraction,
 and  $dN_{\pi\pi}/N_{\pi\pi} dt$ is the normalized
invariant mass spectrum of the two-pion final state. 

The expression (\ref{eq:FFtau}) includes all the corrections which ensure that $F^-(t)$ is the proper quantity to be used in the evaluation (\ref{eq:amu}) of $a_\mu$.   $S_{\rm EW}$ is a short distance correction 
to the effective four-fermion coupling $\tau^-\rightarrow\nu_\tau(d\bar{u})^-$ and  $G_{\rm EM}$ 
is a long distance radiative correction  involving real and virtual photons, calculated in \cite{FloresBaez:2006gf}
for the energy region of interest in our work. 
The isospin breaking due to the mass difference between 
charged and neutral pions is introduced through the  phase-space
\begin{align}\label{eq:beta}
 \beta_{-}(t)\!=\!\left(\!1-\frac{(m_{\pi^-}+m_{\pi^0})^2}{t}\!\right)^{1/2}\! \left(\!1-\frac{(m_{\pi^-}\!- m_{\pi^0})^2}{t}\!\right)^{1/2}\!,\end{align}
relevant in $\tau$ decay. 
We have used standard values \cite{PDG} for the masses  and the CKM matrix element  $|V_{\rm ud}|$.  For
$S_{\rm EW}$ we have used the values given by the experiments themselves.

Several other corrections considered in the literature are small and can be neglected in  the region of interest, 0.65-0.71 GeV. The contribution due to the charged and neutral $\rho$ mass difference  is negligible.
The up and down quark mass difference accounting for 
charge-changing hadronic current between $u$ and $d$ quarks, which
in turn leads to a breakdown of the CVC
hypothesis, introduces a correction of the order of 
$(m_u-m_d)^2/m_\tau^2\simeq 10^{-5}$ \cite{etapi1} to $\cal{B}_{\pi\pi}$. 
Another correction suggested recently for $\tau$ data  is produced by the $\rho-\gamma$ mixing \cite{Jegerlehner:2011ti}. The evaluation in a field-theoretic approach  in \cite{Jegerlehner:2011ti} shows that the effect is important especially above the $\rho$ peak.  We have studied the mixing thoroughly using the 
formulas given in \cite{Jegerlehner:2011ti}  
and did not find an appreciable effect at the lower energies we are interested in.
We have therefore not included the effect.


\end{document}